\documentclass[amsmath, amssymb, preprintnumbers, showpacs, showkeys, aps,prb,superscriptaddress,twocolumn]{revtex4-1}
\usepackage{graphicx}
\usepackage{braket}
\usepackage{ulem}   % to strike things out
\usepackage{dsfont}
\usepackage{amsthm,amsmath,amsfonts,amssymb,verbatim,color}
\usepackage{mathtools} % for \coloneqq
\usepackage{bbold}
\usepackage{graphicx}
\usepackage[T1]{fontenc}
\usepackage[colorlinks=true,citecolor=blue,linkcolor=blue,urlcolor=blue]{hyperref}
\newcommand{\bs}[1]{{\boldsymbol{#1}}}

\newcommand{\tr}{\mathop{\mathrm{Tr}}}

\renewcommand{\b}[1]{{\mathbf{#1}}}

\begin{document}
\preprint{}

\title{Modular Arithmetic with Nodal Lines: Drumhead Surface States in \mbox{ZrSiTe}}

\author{Lukas Muechler}
\affiliation{Center for Computational Quantum Physics, The Flatiron Institute, New York, New York, 10010, USA}
\author{Andreas Topp}
% \email{a.topp@fkf.mpg.de}
\affiliation{Max-Planck-Institut f\"ur Festk\"orperforschung, Heisenbergstrasse 1, D-70569 Stuttgart, Germany}
\author{Raquel Queiroz}
% \email{raquel.queiroz@weizmann.ac.il}
\affiliation{Department of Condensed Matter Physics, Weizmann Institute of Science, Rehovot 76100, Israel}
%\affiliation{Max-Planck-Institut f\"ur Festk\"orperforschung, Heisenbergstrasse 1, D-70569 Stuttgart, Germany}
\author{Maxim Krivenkov}
\affiliation{Helmholtz-Zentrum Berlin f\"ur Materialien und Energie, Elektronenspeicherring BESSY II, Albert-Einstein-Stra\ss e 15, 12489 Berlin, Germany}
\author{Andrei Varykhalov}
\affiliation{Helmholtz-Zentrum Berlin f\"ur Materialien und Energie, Elektronenspeicherring BESSY II, Albert-Einstein-Stra\ss e 15, 12489 Berlin, Germany}
%\author{Bettina V. Lotsch}
%\affiliation{Max-Planck-Institut f\"ur Festk\"orperforschung, Heisenbergstrasse 1, D-70569 Stuttgart, Germany}
\author{Jennifer Cano}
\affiliation{Center for Computational Quantum Physics, The Flatiron Institute, New York, New York, 10010, USA}
\affiliation{Department of Physics and Astronomy, Stony Brook University, Stony Brook, New York 11974, USA}

\author{Christian R. Ast}
\affiliation{Max-Planck-Institut f\"ur Festk\"orperforschung, Heisenbergstrasse 1, D-70569 Stuttgart, Germany}
\author{Leslie M. Schoop}
\affiliation{Department of Chemistry, Princeton University, Princeton, New Jersey 08544, USA}

\begin{abstract}
We study the electronic structure of the nodal line semimetal \mbox{ZrSiTe} both experimentally and theoretically.
We find two different surface states in ZrSiTe - topological drumhead surface states and trivial floating band surface states.
Using the spectra of Wilson loops, we show that a non-trivial Berry phase that exists in a confined region within the Brillouin Zone gives rise to the topological drumhead-type surface states. 
The $\mathbb{Z}_2$ structure of the Berry phase induces a $\mathbb{Z}_2$  'modular arithmetic' of the surface states, allowing surface states deriving from different nodal lines to hybridize and gap out, which can be probed by a set of Wilson loops. 
Our findings are confirmed by \textit{ab-initio} calculations and angle-resolved photoemission experiments, which are in excellent agreement with each other and the topological analysis.
This is the first complete characterization of topological surface states in the family of square-net based nodal line semimetals and thus fundamentally increases the understanding of the topological nature of this growing class of topological semimetals.
\end{abstract}
\date\today
\maketitle

\section{Introduction}
Band inversions in three-dimensional (3D) materials can lead to a variety of topological semimetals that can be distinguished by the dimensionality and connectivity of the band touching points.\cite{yan2017topological,armitage2018weyl,yang2018symmetry,schoop2018chemical,klemenz2019topological}
For example, Weyl semimetals are characterized by isolated points in the Brillouin zone (BZ) at which two bands cross; such crossings are protected by only translation symmetry. 
Protected crossings of a larger number of bands at isolated points within the BZ require the presence of additional spatial symmetries, e.g. Dirac semimetals with four-fold degenerate crossing points require the presence of rotation symmetries \cite{yang2014classification,gibson2015three}.
In materials with nonsymmorphic symmetries or in certain magnetic space groups, band crossings of up to eight bands can be found at high-symmetry points, which is the theoretical maximum \cite{bradlyn2016beyond,Wieder2016Double,schoop2018tunable,cano2019multifold}.
In the presence of mirror symmetries, band inversions can lead to the existence of nodal lines, i.e. one-dimensional (1D) lines or loops of either two- or four-fold degenerate band touching points in the BZ \cite{kim2015dirac,chan2015topological,yamakage2015line,SchnyderNodal}. 
If multiple nodal lines are present in one material, the band structure can further be characterized by their connectivity or linking structure, e.g. it is possible for nodal lines to form knotted nodal structures that are characterized by knot invariants~\cite{bzduvsek2016nodal,bi2017nodal,ezawa2017topological}. \\
These band crossings often strongly influence the electronic properties of such topological semimetals.~\cite{hu2019transport}
For example, the presence of Weyl nodes in the band structure leads to the existence of Fermi arcs~\cite{lv2015experimental,yang2015weyl,huang2015weyl} on the surface of the material or to a measurable transport signature in the longitudinal magnetoresistance, the so-called chiral anomaly \cite{huang2015observation,arnold2016negative,zhang2016signatures,hirschberger2016chiral,liang2018experimental}.
In addition, optical properties of topological semimetals are directly related to the topological invariants protecting the band crossings, such as a quantized circular photogalvanic effect.~\cite{de2017quantized,ma2017direct} \\
In recent years, topological nodal line semimetals have become an active field of research~\cite{yu2017topological,klemenz2019topological}.
Nodal lines are characterized by local topological invariants~\cite{SchnyderNodal}, which guarantee the presence of topologically protected drumhead surface states that can be measured in angle-resolved photoemission spectroscopy (ARPES) experiments~\cite{bian2016topological,belopolski2017three,Liu2018Experimental}.
In addition, nodal lines can serve as important sources of Berry curvature or spin-Berry curvature, which contribute to observables that can be probed in transport experiments, such as the anomalous-Hall effect or the spin-Hall effect \cite{manna2018heusler}. 
For example, it has recently been shown that the giant anomalous Hall effects in magnetic Weyl semimetals such as \mbox{Co$_2$MnGa} and \mbox{Co$_3$Sn$_2$S$_2$} derives from the Berry curvature around nodal lines \cite{AHEHeusler,liu2018giant}.
Similarly, large spin-Hall effects, based on the presence of nodal lines, have been predicted in non-magnetic compounds such as \mbox{RuO$_2$}, \mbox{TaAs} or \mbox{W$_3$Ta} \cite{spinhall1,spinhall2,spinhall3}.
Furthermore, novel transport properties such as unconventional mass enhancement or electron-hole tunneling, were discovered in nodal line semimetals \cite{pezzini2018unconventional,van2018electron}. \\
It has recently been established that the presence of nodal lines in the band structure is related to certain structural motives of the crystal structure \cite{chan20163,klemenz2019topological}.
For example, the family of $MXZ$ ($M$ = Zr,Hf; $X$ = Si,Ge; $Z$ = S,Se,Te) materials exhibits multiple nodal lines and nonsymmorphic degeneracies right at the Fermi level due to the presence of a two-dimensional square net of $X$-atoms in its crystal structure~\cite{schoop2016dirac,topp2016non}.
While much attention has been directed towards the bulk properties of this class of materials, the topological drumhead surface states that are expected to derive from the presence of the nodal lines in the band structure have not been discussed in detail so far~\cite{Nakamura2019Evidence,fu2019dirac} and a general theoretical understanding is lacking.
In this paper, we analyze the nodal line structure of \mbox{ZrSiTe} as a representative member of this class of materials and demonstrate the existence of topologically required surface states both theoretically and experimentally.
We show that the surface states exhibit a $\mathbb{Z}_2$ modular arithmetic according to their $\mathbb{Z}_2$ quantized Berry phase, which can be probed by a set of bulk Wilson loops. ARPES data confirms the existence of these surface states in the areas of non-trivial Berry phase experimentally.\\

The paper is organized as follows: In the first section we review the electronic structure of \mbox{ZrSiTe} and  compare it to its close relative \mbox{ZrSiS} with a particular emphasis on the nodal line structure.
We then move to discuss the topological properties of \mbox{ZrSiTe} based on a Wilson loop analysis and discuss the implications for the surface states of the (001) surface.
In the second section we discuss angle-resolved photoemission spectroscopy (ARPES) measurements on the (001) surface of \mbox{ZrSiTe} and compare it with the theoretical predictions.

\section{Methods}
\subsection{Theoretical}
The DFT calculations were performed using the VASP package~\cite{VASP} with the standard pseudopotentials for Zr, Si and Te. The experimental geometries were taken from the ICSD. For the self-consistent calculations, the reducible BZ was sampled by a $7\times7\times5$ k-mesh. A Wannier interpolation using 82 bands was performed by projecting onto an atomic-orbital basis centered at the atomic positions, consisting of Zr 5$s$,6$s$,5$p$,4$d$,5$d$, Si 3$s$,4$s$,3$p$,4$p$,3$d$ as well as Te 5$s$,6$s$,5$p$,6$p$,5$d$ orbitals.
The nodal-lines and Wilson loops were calculated with an in-house code and the \textit{wanniertools}~\cite{wu2018wanniertools} package.

\subsection{Experimental}
The synthesis and characterization of ZrSiTe single crystals was published elsewhere \cite{topp2016non}.
ARPES experiments were performed on \textit{in situ} cleaved crystals in ultrahigh vacuum (low 10$^{-10}$\,mbar). The spectra were recorded at 50\,K with the 1$^2$ ARPES experiment installed at the UE112-PGM2a beam line at the BESSY-II synchrotron.

\section{Electronic structure}
\label{sec:ElStruc}
\begin{figure}[t]
 % \centering
 \includegraphics[width=0.95\columnwidth]{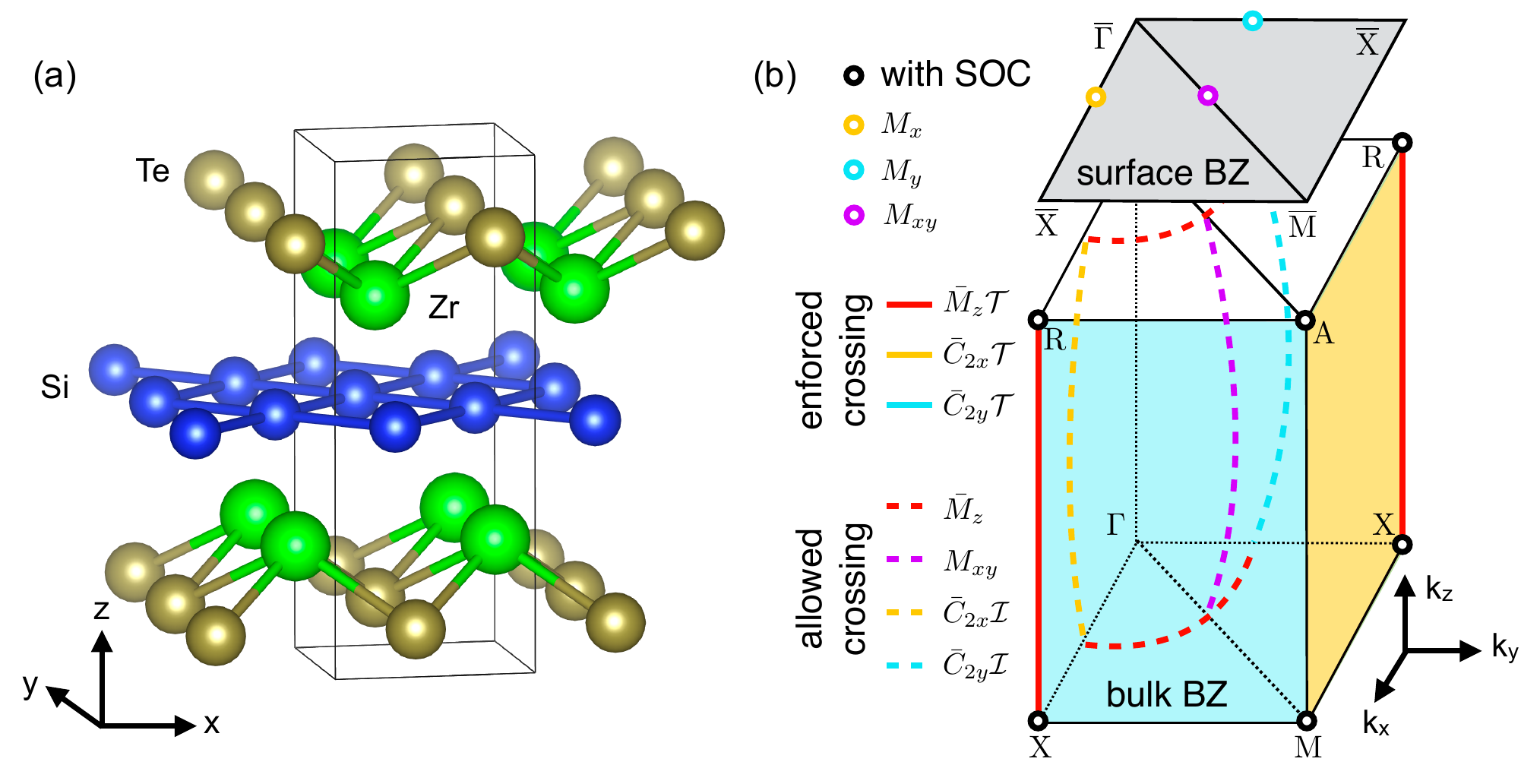}
\caption{
(a) Crystal structure of \mbox{ZrSiTe}, Si atoms are displayed in blue. Zr in green and Te in brown. 
(b) Brillouin zone and crossings enforced/allowed by the space-group symmetries of $P4/nmm$.
}
\label{fig_struc}
\end{figure}

\subsection{Nodal lines}
\begin{figure*}[t]
 % \centering
 \includegraphics[width=1.9\columnwidth]{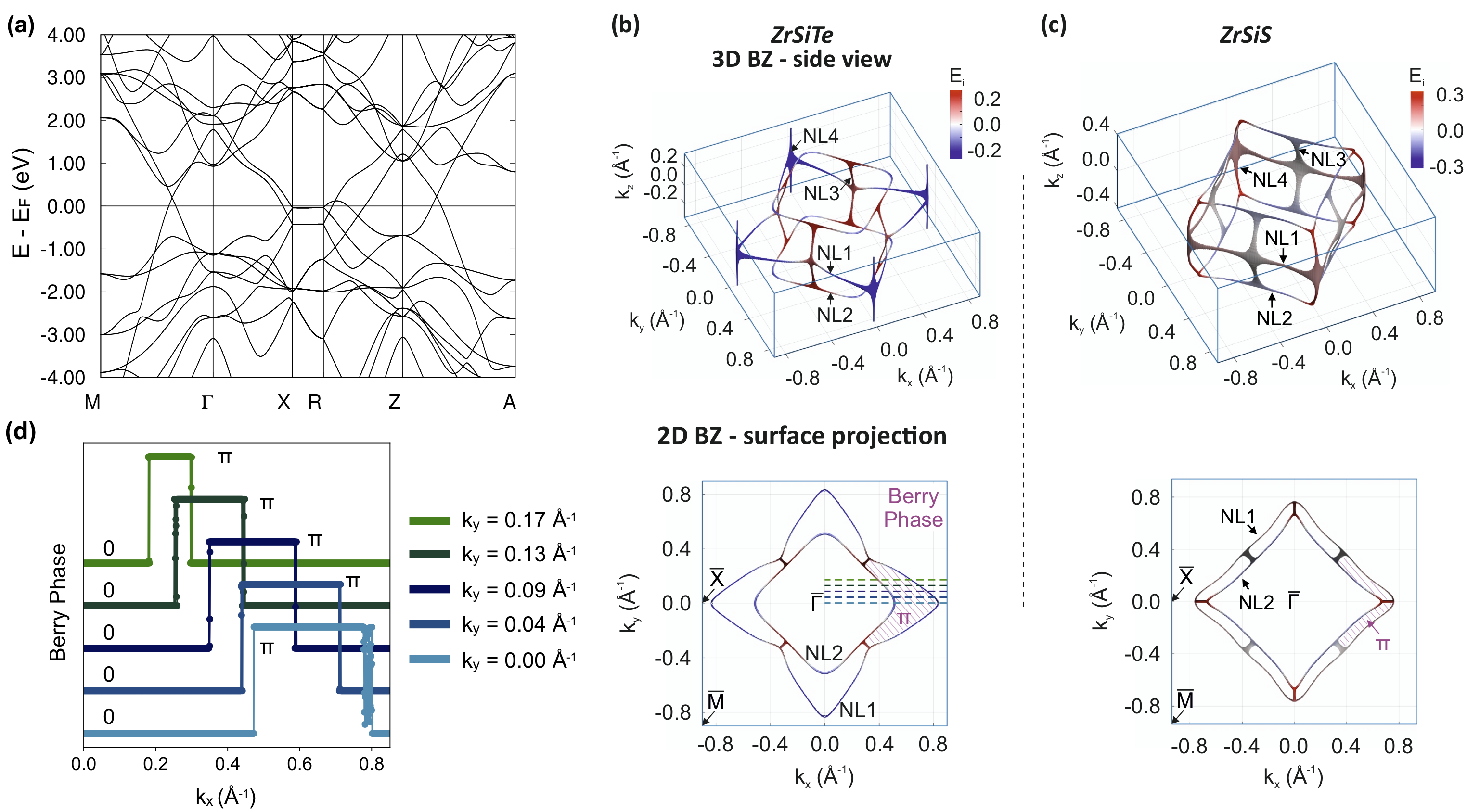}
\caption{
 (a) Bulk band structure of \mbox{ZrSiTe} without SOC. (b) \mbox{ZrSiTe} nodal line connectivity in the 3D BZ (top) and (001) surface projection (bottom). The area of calculated $\pi$ Berry phase is indicated in purple in one quadrant.
(c)  Nodal line connectivity of \mbox{ZrSiS} analogous to (b). 
(d) Berry phase calculated from Wilson loops along $k_z$ as a function of $k_x$ for fixed $k_y$. A Berry phase of $\pi$ implies the presence of topological drumhead surface states at the (001) surface. The constant $k_y$ cuts correspond to the dashed lines in (b).
}
\label{fig_2}
\end{figure*}
Just as its close relative \mbox{ZrSiS}, \mbox{ZrSiTe} crystallizes in the nonsymmorphic space group $P4/nmm$ (SG 129) [Fig.\,\ref{fig_struc}(a)].
Symmetries are key towards the understanding of the electronic structure of  materials, as they protect the crossings of bands in the Brillouin Zone (BZ) in high-symmetry planes, lines or points.
In the case of \mbox{ZrSiS} and \mbox{ZrSiTe}, the following symmetries are of particular importance: the glide mirror $\bar{M}_z = \{ M_z | \frac{1}{2}\frac{1}{2}0\}$, the mirror $M_{xy}$, spatial inversion $\{I | 0 0 0\}$ and the two screw symmetries $\bar{C}_{2x} = \{C_{2x} | \frac{1}{2} 0 0 \}$ and $\bar{C}_{2y} = \{C_{2y} |0 \frac{1}{2} 0 \}$.
In combination with time-reversal symmetry $\mathcal{T}$, these symmetries lead to nonsymmorphically enforced degeneracies at the BZ-zone boundary and the occurrence of multiple nodal lines.
We start our analysis without accounting for spin-orbit coupling (SOC); the presence of SOC will be discussed in detail later.
The combination of a screw axes with time-reversal symmetry $\mathcal{T}$ enforces double degenerate states (ignoring spin) in the $\mathrm{M}$-$\mathrm{X}$-$\mathrm{A}$-$\mathrm{R}$ plane as shown in Fig.\,\ref{fig_struc}(b).
Inside the BZ, $\bar{M}_z$ can protect band crossings in the $k_z = 0,\pi$ plane, while $M_{xy}$ can protect band crossings in the plane given by $k_x = k_y$. The combinations of the screw symmetries with inversion, $\bar{C}_{2x}I$ and $\bar{C}_{2y}I$, allow for crossings in the $k_y = 0$ and $k_x = 0$ planes, respectively.
Due to these many high-symmetry planes, the band structure of \mbox{ZrSiTe} is complex with a plethora of band crossings [Fig.\,\ref{fig_2}(a)]. For this reason, we only focus on crossings close to the Fermi level in the following discussion.
In the vicinity of the Fermi level, two nodal lines protected by $\bar{M}_z$ can be found in the $k_z = 0$ (NL1) and $k_z = \pi$ plane (NL2), respectively.
The dispersions of NL1 and NL2 are noticeably different from each other [Fig.\,\ref{fig_2}(b)], indicating that the electronic structure is of 3D character, despite \mbox{ZrSiTe's} layered crystal structure and the fact that bulk crystals can be easily exfoliated~\cite{hu2016evidence}.
The two nodal lines are connected by an additional nodal line (NL3) in the perpendicular direction, which is protected by $M_{xy}$.
Furthermore, a nodal line (NL4), which is protected by $\bar{C}_{2x}$, exists along the $k_z$-direction that connects to the nodal line in the $k_z = 0$ plane, but does not terminate at the nodal line in the $k_z = \pi$ plane.
The resulting cage-like nodal structure is displayed in Fig.\,\ref{fig_2}(b).
In contrast, in \mbox{ZrSiS}, whose nodal line structure is reproduced in Fig.\,\ref{fig_2}(c), the nodal lines that lie in the $k_z = 0$ and $k_z = \pi$ plane (NL 1 and 2) are connected by two additional nodal lines, one protected by $M_{xy}$ and one by $\bar{C}_{2x}I$. 
In summary: in ZrSiS, both NL3 and NL4 connect NL1 and NL2, while in \mbox{ZrSiTe}, only NL3 connects NL1
and NL2.
% One reason for the difference in nodal line connectivity might be related to the difference in $c/a$ ratio between the two compounds (\mbox{ZrSiTe} with $c/a = 2.5718$ \cite{bensch1994structure} and \mbox{ZrSiS} with $c/a = 2.273$ \cite{klein1964zirconium}), which affects the hopping along the $z$-direction.
NL1 and NL2, which have similar shape in \mbox{ZrSiS}, differ significantly in \mbox{ZrSiTe}. This difference is clearly visible in the (001) surface projection of the nodal lines, as shown in the lower part of Fig.\,\ref{fig_2}(b) and (c), and has important consequences for the presence of topological drumhead surface states.

\subsection{Wilson loops and Berry phase}\label{wilson}
Topological surface states are expected to appear in areas of the surface BZ in which the Berry phase $\gamma$ equals $\pi$. The Berry phase can be computed by a Wilson-loop directed in parallel to the surface normal vector.
The Wilson loop is defined as the path-ordered exponential of the non-abelian Berry connection $\b{A}(\boldsymbol{k})_{ij} = \langle {u_{i,\boldsymbol{k}}} |{\nabla_{\boldsymbol{k}}|u_{j,\boldsymbol{k}}} \rangle$~\cite{alexandradinata2016topological,ArisWilson,PRXWTe2}: 

\begin{align} \label{wloopdifferentiable}
\mathcal{W}(\ell) = \overline{\text{exp}}\,\left[{-\int_{\ell}} d\bs{k} \cdot \boldsymbol{A}(\boldsymbol{k})\,\right],
\end{align}
where $\ket{u_{j,\boldsymbol{k}}}$ is an occupied eigenstate of the Hamiltonian and $\ell$ is a path in the BZ with a finite gap between the highest occupied state and the lowest unoccupied one.
The Berry phase $\gamma \in [0,2\pi)$ is determined by the abelian part ($\tr\b{A}$) of the Berry connection and can be obtained from the determinant of the Wilson loop via
\begin{align} \label{BerryDef}
e^{i \gamma} \coloneqq \text{Det}\big[\,\mathcal{W}(\ell) \,\big] = \text{exp}\,\left[{-\int_{\ell}} d\bs{k} \cdot \tr\b{A}(\boldsymbol{k})\,\right].
\end{align}
It is straightforward  to show that quantization of the Berry phase to either $0$ or $\pi$ occurs if the path of the Wilson loop $(\ell)$ is reversed by a unitary symmetry $\bs{g}$ of the Hamiltonian, i.e. \mbox{$\bs{g}: \ell \mapsto -\ell$}, as we now derive.
Under this symmetry the Wilson loop transforms as
\begin{align}
\bs{g}\mathcal{W}(\ell)\bs{g}^{-1}  = \mathcal{W}(\bs{g}\ell)  = \mathcal{W}(-\ell).
\end{align}
Making use of the fact that $\mathcal{W}(-\ell) = \mathcal{W}^{-1}(\ell)$, we arrive at 
\begin{equation}
\begin{split}
\text{Det}\big[\,\bs{g}\mathcal{W}(\ell)\bs{g}^{-1}\,\big] &= \text{Det}\big[\,\mathcal{W}(-\ell)\,\big] \\ 
\implies \text{Det}\big[\,\mathcal{W}(\ell)\,\big]^{\phantom{2}}  & = \text{Det}\big[\,\mathcal{W}(\ell)^{-1}\,\big] \\ 
\implies \text{Det}\big[\,\mathcal{W}(\ell)\,\big]^2  & = 1,
\end{split}
\end{equation}
which implies that $\gamma = 0$  or $ \gamma = \pi$. \\

We are interested in the topological surface states of the (001) surface of ZrSiTe,
which we analyze by  Wilson loops along $k_z$ starting from $k_z = -\pi$ and ending at $k_z = \pi$, while $k_x$ and $k_y$ remain constant.
In this case, the Berry phase becomes a function of the base points $(k_x,k_y)$, i.e. $\gamma \equiv \gamma(k_x,k_y)$, while $\bar{M}_z$ ensures the quantization of the Berry phase to either $0$ or $\pi$.
In areas of the surface BZ, in which the surface projections of the nodal lines NL1 and NL2 overlap, the total Berry phase is expected to be $\gamma = 2\pi$, since each nodal line contributes $\pi$ to the abelian part of the Wilson loop along $k_z$.
However, since the Berry phase is defined modulo $2\pi$, no topological surface states are expected in regions where the nodal lines project on top of each other.
On the other hand, in regions of the surface where only
one nodal line projects, the Berry phase will be quantized to the nontrivial value of $\pi$ and drumhead surface
states are expected.
Due to the larger momentum separation in the surface projection between NL1 and NL2 in \mbox{ZrSiTe} compared to \mbox{ZrSiS}, the area in which drumhead states can be observed is significantly larger for \mbox{ZrSiTe} as shown in Fig.~\ref{fig_2} (a) and (b). 
Fig.\,\ref{fig_2}(d) shows the Berry phase calculated as a function of $k_x$ for a set of fixed values of $k_y$ for all occupied bands.
For small values of $k_x$, as long as the base-point is inside the overlap region of the nodal lines NL1 and NL2, the calculated Berry phase is equal to 0. As shown in Figure \ref{fig_2}(d), the Berry phase adopts a value of $\pi$ in areas only one nodal line projects to, while it is zero elsewhere. The Berry phase abruptly changes between 0 and $\pi$ upon crossing a nodal line boundary. 
% Therefore, inside NL2 (around $\Gamma$), the Berry phase is zero, since a boundary was crossed twice. Thus, a nontrivial Berry phase only remains in areas, where just one nodal line contributes to the Fermi surface.
%As $k_x$ is increased while $k_y$ remains constant, the Berry phase sharply increases from $0$ to $\pi$.
%The jump occurs at values of $(k_x,k_y)$ that belong to NL2 (located in the $k_z = \pi$ plane), since is ill-defined at gap closing points. 
%After crossing NL2, the Berry phase becomes quantized to $\pi$, which implies the presence of topological surface states. 
%Upon further increase of $k_x$, the Berry phase jumps back to $0$ again, due to crossing a set of $(k_x,k_y)$ which belong to NL1 (located in the $k_z = 0$ plane).
The total area in which the Berry phase equals $\pi$ is highlighted by dashed violet lines in panel (b). 

\begin{figure}
 % \centering
 \includegraphics[width=1.0\columnwidth]{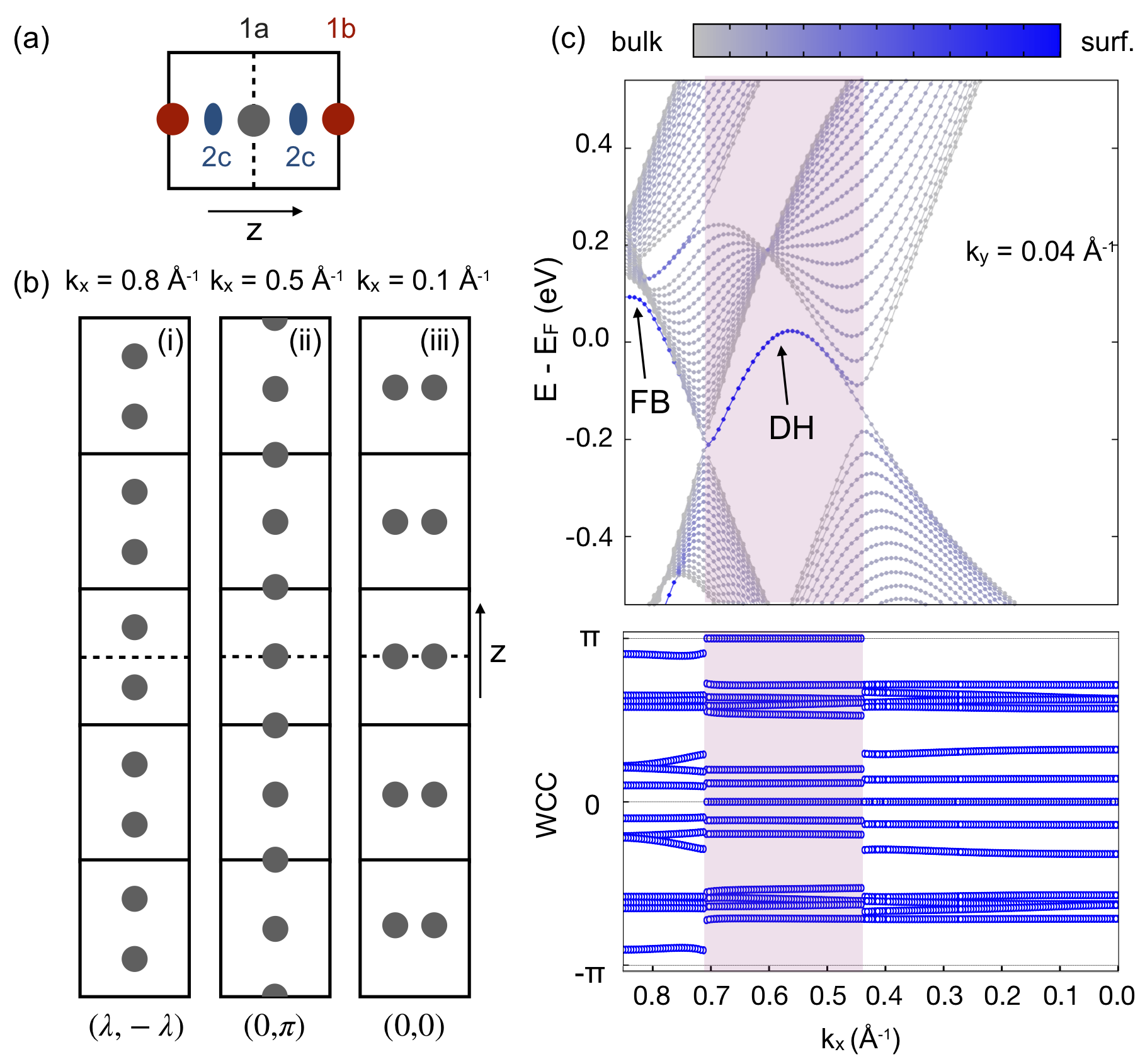}
\caption{
(a) WCC positions constrained by $\bar{M}_z$ in a single unit cell.
(b) Three different $\bar{M}_z$ symmetric slabs with different configurations of WCCs corresponding to the three different configurations found in ZrSiTe [compare to panel (c)]:
(i) WCC located at Pos.\,$2c$, (ii) one WCC located at Pos.\,$1a$ and one at $1b$ (iii) two WCCs located at Pos.\,$1a$. Only two WCCs are shown for the sake of clarity.
(c) Surface band structure calculated from a Wannier interpolation and WCCs calculated along the same $k$-path. Bulk states are depicted in gray, while surface states are colored blue. Drumhead (DH) surface  states in regions with a $\pi$-Berry phase (shaded region) are of topological origin, while the other surface states (FB) close to $E_F$ derive from the local breaking of the nonsymmorphic symmetry $\bar{M}_z$ at the surface~\cite{topp2017surface}. 
}
\label{fig_surf_wilson}
\end{figure}

So far we have only discussed the abelian part of the Wilson loop and its relation to the surface states. The non-abelian part, i.e. the eigenvalues of the Wilson loop, allow to make more detailed statements of the topological structure, in particular the interplay between space-group symmetries, the nodal lines, and their surface states~\cite{ArisWilson,PRXWTe2}.
The unimodular eigenvalues $e^{i\phi_i}$ of the Wilson loop, whose phases $\phi \equiv \phi_i(k_x,k_y)$ depend on the base point of the Wilson loop, are called Wannier charge centers (WCCs) or non-abelian Berry phases.
For a Wilson loop along the $k_z$ direction, $\phi_i(k_x,k_y)$ is related to the charge density of a Wannier function $w_i$ that is maximally localized in the $z$-direction.
% \begin{equation}
    % \phi_i(k_x,k_y) = \frac{1}{2\pi} \int_{UC} dz \braket{w_i|\hat{z}|w_i},
% \end{equation}
The WCC can thus be interpreted as a point like charge density defined within a single unit cell for a fixed base point $(k_x,k_y)$~\cite{Zak1982Band,Zak1989Berry}, effectively mapping each base point to a one-dimensional problem analogous to the SSH-model~\cite{Su1979Solitons}.
The sum of the WCCs is equal to the Berry phase, i.e.
\begin{equation}
    \gamma = \sum_i \phi_i, 
\end{equation}
and is related to the electric polarization in the unit cell via $P = \frac{e}{2\pi} \gamma \ \text{mod }e $, where $e$ is the electric quantum of charge~\cite{King1993Theory}.\\
The positions of the WCCs are constrained by space-group symmetries~\cite{ArisWilson,PRXWTe2}.
In the case of ZrSiTe, the nonsymmorphic mirror $\bar{M}_z$ constrains the positions of the WCCs to three locations [Fig.\,\ref{fig_surf_wilson}(a)]:
(i) Two positions with multiplicity one, located either at the location of the mirror plane at the origin (Pos.\,$1a$), corresponding to $\phi(k_x,k_y) = 0$, or the unit-cell boundary (Pos.\,$1b$), corresponding to $\phi(k_x,k_y) = \pi$.
(ii) One position with multiplicity two (Pos.\,$2c$), corresponding to two WCCs $\phi_1(k_x,k_y),\phi_2(k_x,k_y)$ located between Pos.\,$1a$ and $1b$, i.e. $\phi_1(k_x,k_y) = - \phi_2(k_x,k_y)$.
An odd number of WCCs located at Pos.\,$1a$ is topologically inequivalent to the same number of WCCs located at Pos.\,$1b$, since a single WCC cannot be moved from $1a$ to $1b$ without breaking the mirror symmetry or going through a gap closing point.\\
% due to the higher multiplicity of Pos.$2c$. \\
The location of the WCC constrains the surface state spectrum.
A surface can be modeled by creating a mirror symmetric slab of unit cells along the $z$ direction~[Fig.\,\ref{fig_surf_wilson}(b)]. Topological surface states occur if the surface termination cuts through an odd number of WCCs; in our case this corresponds to an odd number of WCCs located at Pos.\,$1b$ and thus to a bulk Berry phase $\gamma = \pi$.
The WCC are completely determined by the $\bar{M}_z$ eigenvalues of the occupied states at $k_z = 0$ and $k_z = \pi$ via an exact mapping that determines the eigenvalues uniquely (see App.\,\ref{app:Wilson})~\cite{ArisWilson}.

\begin{table}[t]
\setlength{\tabcolsep}{0.5em} 
\centering
{\renewcommand{\arraystretch}{2.1}
\begin{tabular}{c | c c c}
\hline \hline
$k_x / \text{\AA}^{-1}$ & 0.1 & 0.6 & 0.8  \\
\hline
$N_{+,k_z = 0}$         & 6   & 6   & 7    \\
$N_{-,k_z = 0}$         & 8   & 8   & 7    \\
$N_{+,k_z = \pi}$       & 6   & 7   & 7    \\
$N_{-,k_z = \pi}$       & 8   & 7   & 7  \\ \hline 
$N_{+1}$         & 2   & 1   & 0  \\ 
$N_{-1}$       & 0   & 1   & 0  \\
$N_{\alpha,\alpha^{\ast}}$       & $2\times 6$   & $2\times 6$   & $2\times 7$  \\ 
\hline \hline
\end{tabular}
}

\caption{Multiplicity $N_{\pm,k_z}$ of the positive and negative branches of the $\bar{M}_z$ eigenvalues in the two mirror invariant planes $k_z =0,\pi$ for three selected $k_x$ at constant \mbox{$k_y = 0.04$\,\AA$^{-1}$.}
In addition, we list the number $N_{\pm1}$ of Wilson-loop eigenvalues quantized to $\pm 1$ and number $N_{\alpha,\alpha^{\ast}}$ of complex-conjugate pairs $\alpha,\alpha^{\ast}$ obtained by the mapping discussed in App.\,\ref{app:Wilson} at each $k_x$. At each k-point, 14 bands are occupied.
}\label{tab:wilson}
\end{table}

To exemplify this mapping and to validate our topological analysis, we show  the band structure of the (001) surface and the corresponding WCCs as a function of $k_x$ for fixed $k_y=0.04\,\text{ \AA}^{-1}$ in Fig.\,\ref{fig_surf_wilson}(c).
The total number of occupied bands at each k-point is 14 and the corresponding symmetry eigenvalues of each occupied band are shown in Tab.\,\ref{tab:wilson}.
Topological drumhead states are clearly visible for $k$-points that lie between the projections of the two nodal lines NL1 and NL2 which is the region where the Berry phase is quantized to $\pi$. 
Therefore, two WCCs are quantized to $0,\pi$ respectively, i.e. both Pos.$\,1a$ and $1b$ are occupied.
The other WCCs come in pairs $(\lambda,-\lambda)$ located on Pos.$\,2c$ [compare Fig.\,\ref{fig_surf_wilson}(b)]. 
In the region that only contains the projection of one nodal line ($0.44 < k_x  < 0.7$), the number of positive and negative $\bar{M}_z$ eigenvalues of the occupied bands  at $k_z = \pi$ is equal, while there are two more positive than negative eigenvalues in the $k_z = 0$ plane.
In the areas containing the projections of NL1 and NL2 ($k_x < 0.44$), both the number of positive and negative $\bar{M}_z$ eigenvalues of the occupied bands at $k_z = 0$ and $k_z = \pi$ differ by two, leading to two WCCs quantized to $0$, while the others come in pairs of $(\lambda,-\lambda)$.
For $k_x > 0.7$, the number of positive and negative $\bar{M}_z$ eigenvalues is equal for both $k_z = 0$ and $k_z = \pi$ and the WCCs occur only in pairs $(\lambda,-\lambda)$.
An additional set of surface states emerges from the bulk bands in this region, despite a vanishing Berry phase. 
This surfaces state has been discussed before in ZrSiS~\cite{topp2017surface} and is called the floating band (FB) surface state. It originates from the local breaking of the nonsymmorphic symmetry $\bar{M}_z$ at the surface and is not of topological origin.

\subsection{Spin-Orbit Coupling}
\begin{figure}%[t]
 % \centering
 \includegraphics[width=1.0\columnwidth]{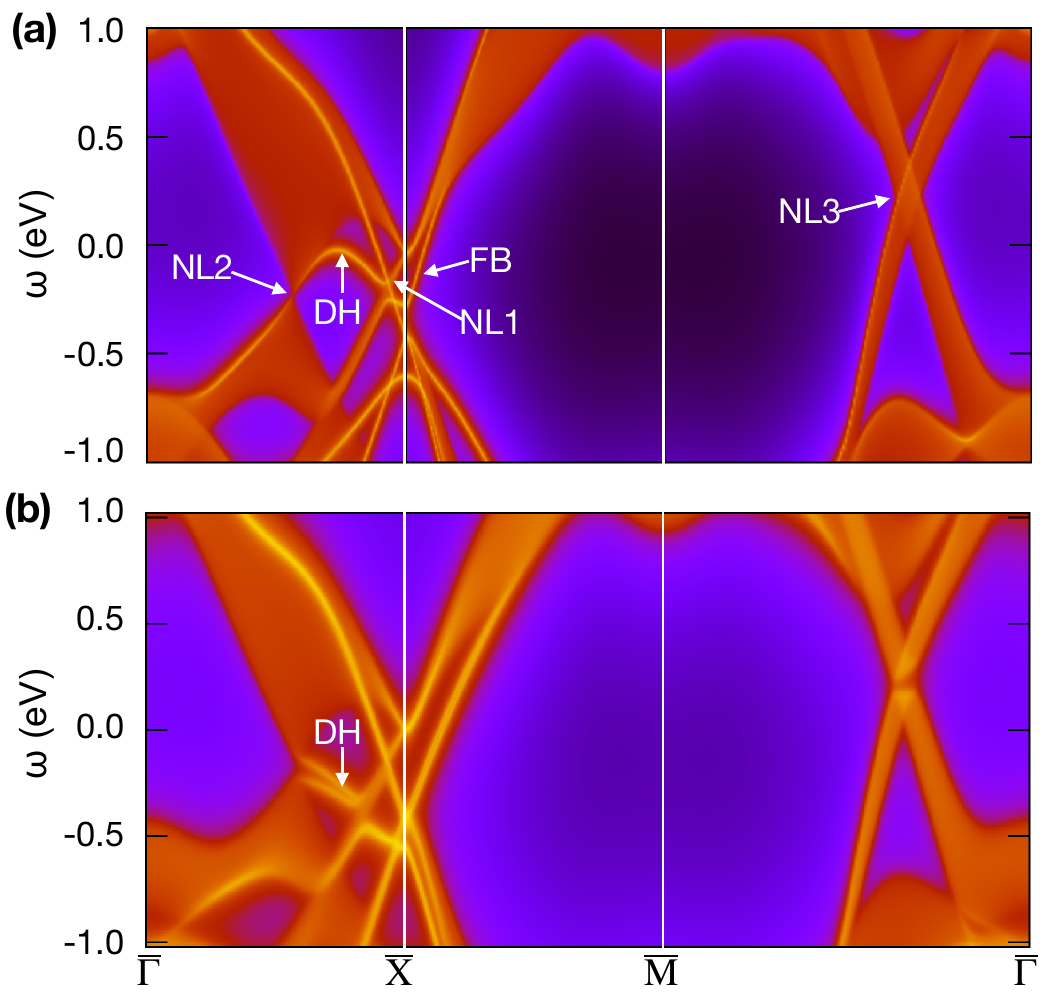}
\caption{
 Surface spectral function of \mbox{ZrSiTe} (001) along surface high-symmetry lines without SOC (a) and with  SOC (b). NL1 and NL2 are the surface projections of the slightly gapped bulk nodal lines in the $k_z = 0$ and $k_z = \pi$ planes.  The drumhead state (DH) emerges from the nodal line NL1 and merges into the bulk states derived from NL2. Along $\mathrm{\overline{X}}$-$\mathrm{\overline{M}}$ a floating band surface state (FB) can be observed, while the projection of the bulk nodal line NL3 that connects NL1 with NL2 can be observed along $\mathrm{\overline{M}}$-$\mathrm{\overline{\Gamma}}$.
}
\label{fig_4}
\end{figure}

It is important to point out that the nodal lines in \mbox{ZrSiTe} are not stable with respect to SOC, i.e. they gap once SOC is considered.
The resulting gaps at the nodal line are of the order of 0.1\,eV in \mbox{ZrSiTe}, which is small compared to the band widths $W \sim 4$\,eV of the bands giving rise to the nodal lines. 
Therefore, we expect that the effects of SOC can be described perturbatively with only slight changes to the topological surface states. 
Our analysis shows that the drumhead states on the (001) surface are two-fold degenerate (counting spin) when SOC is not considered; they split into two branches upon consideration of SOC. 
For weak SOC, the splitting is expected to be small (on the order of a few meV according to our DFT calculations), but is still expected to be observable experimentally.
To verify this hypothesis, we calculated the surface spectral function of \mbox{ZrSiTe} with and without SOC (Fig.\,\ref{fig_4}).
The projections of the bulk nodal lines NL1 and NL2 are clearly visible along the $\mathrm{\overline{\Gamma}}$-$\mathrm{\overline{X}}$ line, while the projection of the nodal line NL3 can be clearly observed along $\mathrm{\overline{M}}$-$\mathrm{\overline{\Gamma}}$.
Similar to the bulk states, the drumhead states are split slightly due to SOC and their dispersion is shifted slightly towards lower energies. 
Yet, they remain as clearly distinguishable features in the surface band structure, which leaves us to conclude that SOC in ZrSiTe can indeed be described as a small perturbation as far as the topological properties are concerned.

\section{ARPES Measurements}
\label{sec:ARPES}

\begin{figure*}[t]
 % \centering
 \includegraphics[width=1.9\columnwidth]{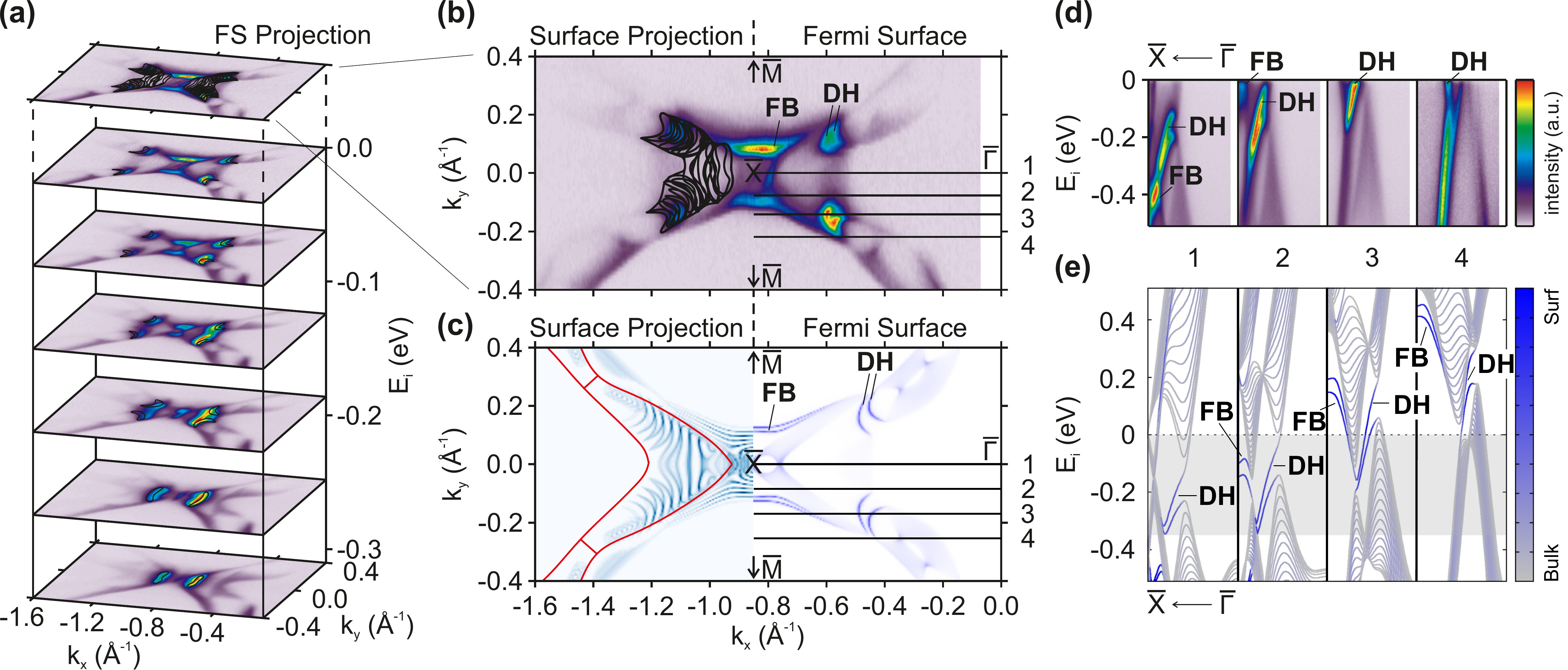}
\caption{
ARPES data, taken at $\hbar\omega=39$\,eV, in comparison with DFT calculations. (a) Constant energy cuts around the $\mathrm{\overline{X}}$ point, tracing the area occupied by the outermost drumhead surface state in black. (b) Magnification of the Fermi surface, showing the surface state projection from (a) on the left. Two different highly intense surface states are labeled as drumhead (DH) and floating band (FB). (c) Slab calculations in analogy to (b). The nodal line is superimposed on the surface projection in red. (d) and (e) Dispersion cuts in experiment and DFT for constant $k_y$ values according to the black lines in (b) and (c), respectively. The FB disperses much stronger, while the DH is located inside the nodal line overlap.}
\label{fig_5}
\end{figure*}

% Since the area in which the drumhead surface states are expected to appear is relatively confined in the overlap area of the nodal lines and ARPES experiments are only sensitive to the 
% occupied part of the band structure, it seems prudent to limit our experimental analysis to the immediate surroundings of the $\mathrm{\overline{X}}$ point, according to the expected surface state dispersion of Fig.\,\ref{fig_4}. 
%The portion of the drumhead states with an energy smaller than $\mathrm{E_F}$ are predicted close to the surface $\mathrm{\overline{X}}$ point.
The topological analysis presented above can be verified with ARPES. Since ARPES experiments are only sensitive to the occupied part of the band structures, we will limit our experimental analysis to the immediate vicinity of the $\mathrm{\overline{X}}$ point, which is where the drumhead states are predicted to appear below $\mathrm{E_F}$.
Fig.\,\ref{fig_5}(a) shows a selection of constant energy cuts from -0.3\,eV to 0\,eV close to the $\mathrm{\overline{X}}$ point. 
At a photon energy of $\hbar\omega = 39\,$eV, the surface state contributions appear as bright features in the band structure, while contributions from the bulk nodal line are only visible with very weak intensity. 
Because of this clear difference in intensity, two crescent-shaped branches of the drumhead surface states, which are split by SOC, can be observed and are labeled in Fig.\,\ref{fig_5}(b). 
%The turning point of the crescent-shaped drumhead states lies very close to the $\mathrm{\overline{X}}$ point. 
The same two crescent shaped drumhead surface states can also be seen in the DFT-calculated Fermi surface shown in Fig.\,\ref{fig_5}(c).
%At the end points of the arc, the drumhead states continue above $E_\mathrm{F}$, which cannot be measured in our experiments. 
To gain an understanding of the total area in which the drumhead states reside and whether this area coincides with that of the nontrivial Berry phase, we traced the area of the drumhead state that lies closest to $\mathrm{\overline{\Gamma}}$, for several initial state energies.
%The area occupied by the outermost surface state in panel (a) is traced in black for several initial state energies.
%Since the drumhead is not flat in energy, i.e. occupies different areas in $k$ space depending on the energy, 
These regions are then superimposed to visualize the overall extent of the drumhead states (black lines in Fig.\,\ref{fig_5}(a) and (b)). %A magnification of this trace can be seen on the left side of Fig.\,\ref{fig_5}(b).
%projected to the Fermi surface and shown on the left side of Fig.\,\ref{fig_5}(b) to create an image of the extent of the drumhead state in the surface BZ. 
The bare measured data is shown without the black surface state projection on the right side of panel (b) as a comparison. This way, the contribution of the drumhead to the Fermi surface can be evaluated directly. In Fig.\,\ref{fig_5}(c), which shows the calculated Fermi surface,
the left side also shows the projection of the surface states between -0.35\,eV and $E_\mathrm{F}$, while the right side presents the bare calculated Fermi surface. On the left side, the bulk nodal line is superimposed as a red line on top of the surface projection. 
The analysis in Fig.\,\ref{fig_5}(b) and (c) shows that, as expected, the drumhead states only exist in the region between NL1 and NL2, which is exactly the region where the Berry phase is nontrivial. %The drumhead surface state (labeled DH) consists of two branches, both having a crescent shape. 
In addition to the drumhead states, another high intensity feature can be observed along the $\mathrm{\overline{X}}$-$\mathrm{\overline{M}}$ direction. 
As mentioned in Sect.\,\ref{wilson}, these states have been previously described as floating bands (labeled FB) \cite{topp2017surface}. They also appear in the calculated Fermi surface in the same region.
These two types of surface states coexist in \mbox{ZrSiTe}, but are of different electronic origin and, therefore, behave very differently in their dispersion plots. 
Such dispersion cuts and their theoretical counterparts are shown in Fig.\,\ref{fig_5}(d) and (e), respectively. The dispersion plots were chosen to show constant $k_y$ values along and in parallel to the high-symmetry line $\mathrm{\overline{\Gamma}}$-$\mathrm{\overline{X}}$, indicated by the black lines in panel (b) and (c). In the calculated band structure, the surface bands are plotted in blue, while the bulk bands are plotted in grey. The grey shaded area represents the energy window chosen for the surface projection of the left side of panel (c), which captures the energy range occupied by the drumhead surface states up to the Fermi level. On the experimental side, the surface states can again be identified by their high intensity. 
In both, the experimental and the theoretical data, the surface states extend down to the lowest initial state energy along the high-symmetry line of cut 1 ($\simeq 0.4 eV$).
%Along the high-symmetry line of cut 1, the two types of surface states can be identified, which extend down to the lowest measured initial state energy. 
The drumhead states are expected to appear at slightly higher energies, while floating band states reside at lower energies, below -0.35\,eV in the vicinity of the $\mathrm{\overline{X}}$ point. In the measured data, the two SOC-split branches of the drumhead state can be clearly resolved connecting the bulk nodal lines, which are visible as dark shadows in the chosen color scale.
%It is clearly visible in the data that the branches of the drumhead states disperse upwards in energy as a function of $k_x$ and connect to the two bulk nodal lines.
%For lower initial state energies and in the vicinity of the $\mathrm{\overline{X}}$ point, the dispersion plot shows an additional surface state, which can be characterized as a floating band state.
Moving away from the high-symmetry line $\overline{\Gamma}$-$\overline{\mathrm{X}}$ (following cuts 2-4), we can observe the drumhead surface state to slowly disperse upwards in energy, while it remains between the surface projection of the bulk nodal line, in agreement with the calculations. In cut 4, the lower branch of the drumhead surface state barely remains in the picture. The floating bands, on the other hand, show a very different behavior. They exhibit a much steeper dispersion along $k_y$ and move above the Fermi level very rapidly; by cut 3 they are no longer visible.

\section{Conclusion}
In conclusion, we have studied the electronic structure of the (001) surface of the nodal line semimetal \mbox{ZrSiTe}, both theoretically and experimentally.
We find that the nodal lines in the $k_z = 0$ and $k_z = \pi$ plane of bulk \mbox{ZrSiTe} give rise to clearly recognizable topological drumhead surface states close to $E_\mathrm{F}$, which remain clearly identifiable if SOC is considered. 
The drumhead states gap in regions where they overlap, leading to a ribbon of drumhead states confined by the surface projections of the bulk nodal lines due to their $\mathbb{Z}_2$ classification.
In addition to these topologically required surface states, we find topologically trivial floating band states close to the surface high-symmetry $\mathrm{\overline{X}}$ point.
These states derive from surface symmetry breaking and have been previously reported in the closely related compound \mbox{ZrSiS}.
The study of such an interplay of topologically trivial and nontrivial surface states is not limited to \mbox{ZrSiTe} and should be observable in other nodal line semimetals.

 \begin{acknowledgments}
We would like to than Yan Sun (CPFS) for providing the Wannier interpolation.
This work was partially supported by NSF through the Princeton Center for Complex Materials, a Materials Research Science and Engineering Center DMR-1420541. The authors gratefully acknowledge the financial support by the Max Planck Society. We thank HZB for the allocation of synchrotron radiation beamtime. This work was additionally supported by the DFG, proposal no.\,SCHO 1730/1-1. The Flatiron Institute is a division of the Simons Foundation.
\end{acknowledgments}

\appendix
\section{Mapping between Wilson loop spectrum and symmetry eigenvalues of occupied bands}\label{app:Wilson}
The unimodular spectrum of a Wilson loop $\mathcal{W}(\ell)$ in conjunction with a unitary symmetry $\bs{g}$ that maps $\bs{g}: \ell \mapsto -\ell$ consists of complex conjugate pairs ($\alpha,\alpha^{\ast}$) and eigenvalues that are quantized to $\pm 1$.
The number of complex conjugate pairs $N_{\alpha,\alpha^{\ast}}$ and the number $N_{\pm 1}$ of eigenvalues at $\pm 1$ are completely determined by the eigenvalues of the symmetry $\bs{g}$ in the space of the occupied bands.

In this appendix, we briefly review the algorithm to determine the spectrum of the Wilson loop and refer the reader to references \citet{ArisWilson,alexandradinata2016topological} and \citet{PRXWTe2} for a more detailed derivation in case of symmorphic and nonsymmorphic symmetries respectively.

The algorithm is applicable to symmetries $\bs{g}$ that leave the base-point $\bs{k}$ of the Wilson loop invariant, while reversing the path $\ell$, which is parametrized by a k-vector $\bs{k}_{\|}$.
For each base-point $\bs{k}$, the symmetry commutes with the Hamiltonian at $\bs{k}_{\|} = - \bs{k}_{\|} \mod \bs{G}$, where $\bs{G}$ is a reciprocal lattice vector, and the eigenstates of the Hamiltonian can be labeled by the eigenvalues of $\bs{g}$.
For example, in \mbox{ZrSiTe} $\bs{g}\equiv\bar{M}_z$, while $\bs{k} = (k_x,k_y)$ and $\bs{k}_{\|} = k_z \hat{\bs{k}}_z$, i.e. the electronic states at $(\bs{k},k_z = 0,\pm \pi)$ can be labeled by their $\bar{M}_z$ eigenvalues.
The algorithm to determine the spectrum of $\mathcal{W}(\ell)$ for a Wilson loop along $\bs{k}_z$ is then given as follows:\\

(i) Determine the set $N = \{ N_{+,0}, N_{-,0}, N_{+,\pi},N_{-,\pi}\}$, where $N_{\pm,k_z = 0,\pi}$ is the number of occupied bands at $(\bs{k},k_z = 0,\pm \pi)$ which belongs to the positive (negative) branch of $\bar{M}_z$ eigenvalues $\pm \exp{(-i\frac{k_x + k_y}{2})}$.\\
(ii) Choose the smallest number of this set. The smallest set might be empty and/or not unique, in which case any choice between the equally small sets is valid. We label the smallest number as $N_{\xi,k_z}$, where $\xi$ labels the branch of $\bar{M}_z$ and $k_z =0,\pi$.
It is useful to define $\bar{k}_z$, where $\bar{k}_z = 0$ if $k_z = \pi$ or $\bar{k}_z = \pi$ if $k_z = 0$\\
(iii) The number $N_{-\xi}$ of $-\xi$ eigenvalues is given as $N_{-\xi} = N_{+,\bar{k}_z} - N_{\xi,\bar{k}_z} $ \\ 
(iv) The number $N_{\xi}$ of $\xi$ eigenvalues is given as $N_{\xi} = N_{-,\bar{k}_z} - N_{\xi,k_z} $ \\
(v) The number $N_{\alpha,\alpha^{\ast}}$ of eigenvalues appearing in complex conjugate pairs is given as $N_{\alpha,\alpha^{\ast}} =2 \times N_{\xi,k_z}$. \\ 
We now illustrate the algorithm with an example, using the first column of Tab.~\ref{tab:wilson}, which we reproduce here for convenience:
\begin{equation}
N = \{ N_{+,0} = 6, N_{-,0} = 8 , N_{+,\pi} = 6 ,N_{-,\pi} = 8\}    
\end{equation}
We choose the smallest integer to be $N_{\xi,k} \equiv N_{+,0} = 6$, therefore $\xi  = +1$, $k = 0$  and $\bar{k}  = \pi $.
We thus arrive at $N_{-1} = N_{+,\pi} - N_{+,0} =  6 - 6 = 0$, while  \mbox{$N_{+1} = N_{-, \pi}  -  N_{+, 0} = 8-6 = 2$} and  $N_{\alpha,\alpha^{\ast}}= 2\times N_{+,0} = 12$.

\bibliography{lit}

\end{document}